\documentclass{article}

\usepackage{spconf,amsmath,amssymb,amsfonts,bm,algorithm,graphicx,booktabs,multirow,enumitem,cite,sidecap,nicematrix,xcolor}
\usepackage[noend]{algorithmic}
\usepackage[hidelinks]{hyperref}
\usepackage[mathlines,switch]{lineno}
\usepackage[table]{xcolor}

\newcommand{\vt}{\bm{v}_{\theta}}
\newcommand{\xt}{\bm{x}}
\newcommand{\Rmask}{\mathcal{P}}
\newcommand{\Gmask}{\mathcal{G}}
\newcommand{\E}{\mathbb{E}}
\newcommand{\R}{\mathbb{R}}

\title{Prompt-Consistency Inference for Zero-Shot Flow-Matching Text-to-Speech Models}

\name{Vasily Zadorozhnyy \quad\quad Can Goksen \quad\quad Kazuhito Koishida \quad\quad Dung Tran}

\address{Applied Sciences Group, Microsoft Corporation, Redmond, USA}

\begin{document}
\ninept
\maketitle

\begin{abstract}
In recent years, flow-matching models have produced significant improvements in zero-shot text-to-speech synthesis. Conditioned on an audio prompt and text, these models learn a velocity field and generate speech by iteratively solving an ODE. During inference, the solver evolves a single state spanning both the prompt and the region to be generated, although only the generated region is ultimately retained. The discarded prompt state, however, still matters; its intermediate values influence generation through the velocity field that couples the two regions. As sampling continues, this state can drift away from the prescribed conditional path, introducing a discrepancy into subsequent generation updates. Unlike the unknown generated trajectory, the prompt path is available in closed form from the reference audio and the initial noise. We exploit this observation with Prompt-Consistency Inference (PCI), a training-free rule that restores the prompt block to its analytic value before each velocity evaluation, while leaving the generated block unchanged. PCI improves speaker similarity and intelligibility across the evaluated flow-matching TTS backbones without additional network evaluations. Our ablation studies further show that PCI keeps post-step prompt discrepancies smaller and that corrections covering the later sampling stages recover much of the observed similarity gain.
\end{abstract}


\section{Introduction}
\label{sec:introduction}

A zero-shot Text-to-Speech (TTS) system aims to synthesize voice speech of an unseen speaker from a single short reference utterance, without any speaker-specific fine-tuning~\cite{wang2023neural,le2023voicebox}. Many recent systems use \emph{Flow Matching} (FM) method, which learns a continuous-time velocity field that transports a simple noise distribution to the data distribution along a prescribed probability path~\cite{lipman2023flow,liu2023rectified,tong2024improving,albergo2023building,lim2024elucidating}. The FM TTS models such as Voicebox~\cite{le2023voicebox}, E2~TTS~\cite{10832320_e2tts}, F5-TTS~\cite{chen-etal-2024-f5tts}, and ZipVoice~\cite{zhu2025zipvoice} achieve high synthesis quality while remaining non-autoregressive and efficient to sample at inference.

Several of these systems are \emph{prompted} through audio infilling. The acoustic features of the reference prompt and the region to be generated occupy two temporal blocks of a single state, and the model is conditioned on the clean prompt together with its transcript and the target text. At inference, both blocks are initialized from a single noise sample $\bm{\epsilon}\sim\mathcal{N}(\bm{0},\bm{I})$, and the learned ODE is integrated jointly over the full state from $t=0$ to $t=1$. Once sampling is complete, only the generated region is kept as synthesized speech, while the prompt region is discarded. The two regions thus share a sampling trajectory, despite serving different roles.

Discarding the prompt output, however, does not remove its role during generation. The velocity field processes the two regions jointly at intermediate times, so the update in the generated region depends on the current state of the prompt region at every sampling step. The clean reference supplied as conditioning remains unchanged, but the evolving prompt state follows the model's predicted velocity. If this state departs from the prompt interpolation prescribed during training, the discrepancy becomes part of the input used to compute subsequent generation updates. The prompt block therefore matters not because of its final output, but because of the intermediate values the network sees along the way.

The prompt region also has a property that the generated region does not: its prescribed conditional path is known in closed form. Under the linear conditional FM construction~\cite{lipman2023flow}, this path is an interpolation between the reference features and the fixed noise drawn at initialization. Since both are available at inference, the deviation of the integrated prompt state from this path can be measured exactly. We call this discrepancy \emph{prompt drift}. It is important to distinguish prompt drift from numerical integration error alone; the predicted prompt velocity need not follow the prescribed interpolation, even if the learned ODE is integrated accurately. What makes prompt drift useful to address is, therefore, not a particular source of error, but the fact that it is observable and can be corrected before it influences the next update.

We introduce \emph{Prompt-Consistency Inference} (\emph{PCI}) to exploit this observation. Before each velocity evaluation, we overwrite the prompt region with its analytic value on the conditional path, leaving the generated region unchanged. To the best of our knowledge, PCI is the first method to explicitly enforce this prompt-path consistency during inference in FM TTS. Rather than asking the model to reconstruct an observed region, PCI supplies its prescribed state directly, while the generated region continues to evolve under the learned field. An ODE solver step can introduce a new prompt residual, but PCI corrects the prompt input before the next velocity evaluation rather than carrying that discrepancy forward. In this way, the known prompt path becomes an explicit constraint on the sampling input. The modification requires no retraining, no additional parameters, and no extra network evaluations for a fixed sampling schedule.

Our contributions can be summarized as follows:
\begin{itemize}[leftmargin=*]
    \item We identify \emph{prompt drift}, an inference-time discrepancy in prompted FM TTS in which the integrated prompt state departs from its known conditional path. We explain why this discrepancy is relevant to generation even if the final prompt output is discarded.
    \item We introduce PCI, a training-free, parameter-free inference rule that restores the prompt state before each velocity evaluation via a simple tensor overwrite. PCI applies to prompted FM models with an observed state region and analytical conditional path.
    \item We evaluate PCI across several TTS backbones and benchmarks, demonstrating improvements in speaker similarity and intelligibility. In addition, our ablation studies examine the post-step discrepancies and show that corrections covering later sampling stages recover much of the observed similarity gain.
\end{itemize}


\section{Related Work}
\label{sec:related_work}

\subsection{Flow Matching and Rectified Flow}
FM~\cite{lipman2023flow} trains a continuous generative model by regressing a time-dependent velocity field onto the conditional velocity of a prescribed probability path, avoiding simulation of the full ODE during training. Rectified flow~\cite{liu2023rectified} and related conditional FM formulations~\cite{tong2024improving} use linear noise-to-data interpolation, providing simple conditional paths with constant velocities. These training paths, however, should be distinguished from the trajectories of the learned field, which need not follow individual noise-data interpolations. PCI exploits the part of the conditional path that remains available at inference: the interpolation between the known prompt features and their initial noise.

\subsection{Prompting and Audio Infilling}
Prompt-based conditioning, popularized for speech by neural codec language models~\cite{wang2023neural} and text-guided audio infilling~\cite{le2023voicebox}, uses a short reference utterance as evidence of speaker identity and prosody. In prompted FM models, the clean reference can serve as conditioning while a noisy version of the prompt also occupies part of the evolving state. These two roles are distinct: keeping the clean reference fixed does not ensure the evolving prompt follows its prescribed path. PCI addresses the second role by constraining the prompt state supplied to the velocity field, without changing the reference conditioning.

\subsection{Flow Matching Text-to-Speech}
Voicebox~\cite{le2023voicebox} casts text-guided speech generation as masked audio infilling, supporting synthesis and editing within one model. Matcha-TTS~\cite{mehta2024matcha} combines an encoder-decoder acoustic model with optimal-transport conditional FM for efficient synthesis. Models such as E2~TTS~\cite{10832320_e2tts} and F5-TTS~\cite{chen-etal-2024-f5tts} simplify explicit alignment components, while ZipVoice~\cite{zhu2025zipvoice} combines a compact backbone with flow distillation to reduce sampling cost. These methods improve model architecture, training, or sampling procedure. PCI instead corrects the evolving prompt input and applies to samplers that jointly represent observed and generated regions with an available conditional prompt path.

\subsection{Guidance and Inference-Time Consistency}
Classifier-Free Guidance (CFG)~\cite{ho2022cfg} controls generation by combining conditional and unconditional predictions. A complementary strategy is to enforce consistency with known observations during sampling. In the image domain, RePaint~\cite{lugmayr2022repaint} replaces known image regions with noised observations in diffusion sampling, while Restora-Flow~\cite{hadzic2026restoraflow} combines time-dependent observation fusion with trajectory correction and resampling in FM restoration. PCI shares the principle of enforcing known information, applying it to conditioned FM TTS through a prompt overwrite using fixed initial noise. Our contribution focuses on identifying, evaluating, and correcting prompt-state inconsistency in the samplers, rather than known-region replacement itself.


\section{Method}\label{sec:method}

\subsection{Background: Conditional Flow Matching}\label{subsec:background}

Let $\bm{f}, \bm{\epsilon}\in\R^{d\times L}$ denote acoustic features and Gaussian noise, respectively, with $d$ channels and $L$ frames, where $\bm{\epsilon}\sim\mathcal{N}(\bm{0},\bm{I})$. Under the linear conditional FM construction~\cite{lipman2023flow,liu2023rectified}, a training state at time $t\in[0,1]$ is obtained by interpolating between noise and data:
\begin{equation}
    \xt_t^{\star} = t\cdot\bm{f}+(1-t)\cdot\bm{\epsilon}.\label{eq:linear_path}
\end{equation}
The corresponding conditional velocity is constant along the path:
\begin{equation}
    \bm{u} = \frac{d\xt_t^{\star}}{dt} = \bm{f}-\bm{\epsilon}.\label{eq:target_velocity}
\end{equation}

During training, we let $\bm{m}\in\{0,1\}^{d\times L}$ be a binary mask selecting the region to be generated, and let $\odot$ denote the Hadamard product. With conditioning $\bm{c}$, the masked regression objective is
\begin{equation}
    \mathcal{L}_{\mathrm{CFM}}  = \E_{\bm{f},\,\bm{m},\,\bm{\epsilon},\,t}
    \left\| \bm{m}\odot\left(\vt(t,\xt_t^{\star};\bm{c})-(\bm{f}-\bm{\epsilon})\right)\right\|_F^2,
    \label{eq:cfm_loss}
\end{equation}
where the expectation is over training examples, masks, noise, and time. The conditioning $\bm{c}$ includes the clean reference features $(\bm{1}-\bm{m})\odot\bm{f}$. Under this masked objective, velocity errors are penalized only in the generated region. The prompt state is therefore supplied as a network input, but its predicted velocity is not directly supervised by this loss. Consequently, using that velocity to evolve the prompt during sampling need not preserve its prescribed interpolation. More generally, the conditional interpolation defines the training input and velocity target; it does not require the learned field to reproduce each individual noise--data path at inference.

To generate speech, we initialize $\xt_{t_0}=\bm{\epsilon}$ and integrate the learned field over a partition $0=t_0<\cdots<t_N=1$. Although other ODE solvers may be used~\cite{10832320_e2tts}, explicit Euler is a common choice~\cite{chen-etal-2024-f5tts,zhu2025zipvoice}, updating the state with step size $h_i=t_{i+1}-t_i$ as
\begin{equation}
    \xt_{t_{i+1}}
    =
    \xt_{t_i}
    +h_i\vt(t_i,\xt_{t_i};\bm{c}).
    \label{eq:euler}
\end{equation}
Unlike the training state $\xt_t^{\star}$, the sampled state $\xt_{t_i}$ depends on successive model predictions evaluated at preceding sampled states. For notational simplicity, $\vt$ in the following equations may include the backbone's existing guidance rule, such as CFG~\cite{ho2022cfg}.

\subsection{Prompted Infilling and Prompt Drift}\label{subsec:prompt_drift}

Let $\Rmask\subset\{1,\ldots,L\}$ index the prompt frames and $\Gmask=\{1,\ldots,L\}\setminus\Rmask$ the frames to be generated. For a prefix prompt, we write the joint state as $\xt=[\,\xt[\Rmask]\mid\xt[\Gmask]\,]$, where $\mid$ denotes temporal concatenation. The conditioning $\bm{c}$ includes the known reference features $\bm{f}[\Rmask]$, the reference transcript, and the text to be generated.

In the baseline joint-state sampler, \eqref{eq:euler} updates both regions, and only $\xt_{t_N}[\Gmask]$ is kept as the generated acoustic features. Although the clean reference in $\bm{c}$ is fixed, the prompt coordinates $\xt_{t_i}[\Rmask]$ evolve from the predicted velocity and need not follow the conditional interpolation path used to construct the training input. The interpolation is available for the prompt region at every sampling time from \eqref{eq:linear_path} using known reference features and fixed initial noise $\bm{\epsilon}$:
\begin{equation}
    \bm{p}_t^{\star} = t\cdot\bm{f}[\Rmask] + (1-t)\cdot\bm{\epsilon}[\Rmask].\label{eq:prompt_path}
\end{equation}
We define \emph{prompt drift} as the difference between the integrated prompt state and this prescribed value:
\begin{equation}
    \bm{\delta}_i = \xt_{t_i}[\Rmask]-\bm{p}_{t_i}^{\star}.\label{eq:drift}
\end{equation}
Since $\xt_{t_0}=\bm{\epsilon}$, the initial prompt drift is $\bm{\delta}_0=\bm{0}$. Furthermore, subtracting the prompt interpolation from the Euler update gives
\begin{equation}
    \bm{\delta}_{i+1} = \bm{\delta}_i + h_i\bigl(\vt(t_i,\xt_{t_i};\bm{c})[\Rmask] - (\bm{f}[\Rmask]-\bm{\epsilon}[\Rmask])\bigr).\label{eq:drift_recursion}
\end{equation}

Equation~\eqref{eq:drift_recursion} shows that each step adds a scaled velocity residual to the prompt drift, which could amplify or offset it. This drift is not from numerical integration error alone: Euler integration of the exact constant conditional velocity would recover the linear prompt path at all sampling nodes, where the learned prompt velocity need not match the target. Any drift that enters the next network evaluation and can affect the generated-region update through the jointly processed state even though the final prompt output is discarded.

\subsection{Prompt-Consistency Inference}\label{subsec:pci}

Since the prompt region path is known, we need not rely on the predicted prompt state when evaluating the next velocity. PCI uses this observation directly. Before each evaluation, it restores the prompt coordinates to \eqref{eq:prompt_path} and the generated coordinates are unchanged. To distinguish the corrected input from the solver output, let $\xt_{t_i}$ denote the raw state produced by the preceding Euler step. PCI constructs the corrected network input and advances the state as follows:
\begin{equation}
    \begin{aligned}
        \xt_{t_i}^{\mathrm{PCI}}
        &= [\,\bm{p}_{t_i}^{\star}\mid\xt_{t_i}[\Gmask]\,],\\
        \xt_{t_{i+1}}
        &= \xt_{t_i}^{\mathrm{PCI}}
           + h_i\vt(t_i,\xt_{t_i}^{\mathrm{PCI}};\bm{c}).
    \end{aligned}
    \label{eq:pci}
\end{equation}
Note that the prompt noise $\bm{\epsilon}[\Rmask]$ is drawn once at initialization and reused throughout sampling, so every correction restores the same conditional path.

By construction, PCI eliminates prompt drift before each velocity evaluation:
$\xt_{t_i}^{\mathrm{PCI}}[\Rmask]-\bm{p}_{t_i}^{\star}=\bm{0}$.
The subsequent Euler update can nevertheless produce a nonzero post-step residual if the predicted prompt velocity differs from the conditional target:
\begin{equation}
    \bm{\delta}_{i+1}^{\mathrm{PCI}}
    =h_i\bigl(
    \vt(t_i,\xt_{t_i}^{\mathrm{PCI}};\bm{c})[\Rmask]
    -(\bm{f}[\Rmask]-\bm{\epsilon}[\Rmask])
    \bigr).
    \label{eq:pci_post_step}
\end{equation}
Algorithm~\ref{alg:pci} presents the full-state implementation with Euler integration, which also allows us to measure the post-step prompt residual in our ablation study (see \S~\ref{subsec:ablation_ot}).

\begin{algorithm}[t]
    \caption{Prompt-Consistency Inference (Euler)}
    \label{alg:pci}
    \begin{algorithmic}[1]
        \REQUIRE Conditioning $\bm{c}$, reference features $\bm{f}[\Rmask]$, block indices $\Rmask,\Gmask$, sampling times $\{t_i\}_{i=0}^{N}$, velocity field $\vt$
        \STATE Sample $\bm{\epsilon}\sim\mathcal{N}(\bm{0},\bm{I})$;
               set $\xt_{t_0}\leftarrow\bm{\epsilon}$
        \FOR{$i=0,\ldots,N-1$}
            \STATE $\xt_{t_i}^{\mathrm{PCI}}[\Gmask]\leftarrow\xt_{t_i}[\Gmask]$
            \STATE $\xt_{t_i}^{\mathrm{PCI}}[\Rmask]\leftarrow
            t_i\bm{f}[\Rmask]+(1-t_i)\bm{\epsilon}[\Rmask]$
            \STATE $\bm{v}\leftarrow
            \vt(t_i,\xt_{t_i}^{\mathrm{PCI}};\bm{c})$
            \STATE $\xt_{t_{i+1}}\leftarrow
            \xt_{t_i}^{\mathrm{PCI}}+(t_{i+1}-t_i)\bm{v}$
        \ENDFOR
        \STATE \textbf{return} $\xt_{t_N}[\Gmask]$
    \end{algorithmic}
\end{algorithm}

\subsection{Remarks}\label{subsec:remarks}

\begin{itemize}[leftmargin=*]
    \item \textbf{Integrator choice.} PCI can also be applied to inputs at intermediate-stage velocity evaluations in higher-order ODE solvers. We evaluate the Euler implementation in this work.

    \item \textbf{Flow-matching path.} PCI does not require a linear interpolation path~\eqref{eq:prompt_path}. It can use other conditional paths, provided the prompt state at each evaluation time can be reconstructed from the known reference and fixed sampled noise consistently with training.

    \item \textbf{Applicability beyond TTS.} PCI is not inherently tied to speech synthesis. The same principle can be applied to other conditional flow-matching models with jointly evolved observed and generated regions, provided the prescribed path on the observed region is available at inference.
\end{itemize}


\section{Experiments and Results}
\label{sec:exp}

\subsection{Experimental Setup}
\label{subsec:setup}

We evaluate PCI on E2~TTS~\cite{10832320_e2tts}, F5-TTS~\cite{chen-etal-2024-f5tts}, and ZipVoice and its distilled variants~\cite{zhu2025zipvoice}, using pretrained checkpoints downloaded from Hugging Face and keeping all model weights frozen. For each model and utterance, baseline and PCI use identical random seeds, initial noise, conditioning, duration settings, ODE solver, sampling schedules, and guidance settings; the only difference is the prompt correction in \eqref{eq:pci}, applied before each velocity evaluation without additional network evaluations. We evaluate on three standard zero-shot benchmarks: LibriSpeech-PC test-clean~\cite{10389666_LibriSpeech_PC}, Seed-TTS test-en~\cite{commonvoice:2020_Seed_EN}, and Seed-TTS test-zh~\cite{Guo2020DidispeechAL_Seed_ZH}.

\subsection{Evaluation Metrics}\label{subsec:metrics}

We report: (i) \textbf{speaker similarity} (SIM-o), the cosine similarity between speaker embeddings of the generated and reference audios, extracted with a WavLM-based~\cite{chen2022wavlm} ECAPA-TDNN speaker-verification model~\cite{desplanques2020ecapa}; (ii) \textbf{word error rate} (WER), computed by comparing the ASR transcription of synthesized speech with the input text, using Whisper-large-v3~\cite{radford2023whisper} for Seed-TTS test-en, Paraformer-zh~\cite{gao2022paraformer} for Seed-TTS test-zh, and a HuBERT-based recognizer~\cite{hsu2021hubert} for LibriSpeech-PC test-clean; and (iii) \textbf{UTMOS}, an automatic prediction of speech naturalness~\cite{saeki2022utmos}. For the trajectory analysis, we additionally report the Frobenius norms of prompt drift and generated-block reference-path discrepancy in \S~\ref{subsec:ablation_ot}.

\subsection{Main Results}\label{subsec:main_result}

Table~\ref{tab:prompt_conditioned_large} summarizes the comparison. PCI improves SIM-o and reduces WER in many settings while improving or preserving UTMOS in most. On Seed-TTS test-en, F5-TTS improves SIM-o from $0.664$ to $0.685$ (relative gain of $3.2\%$). On LibriSpeech-PC, ZipVoice at 16 NFEs reduces WER from $1.77\%$ to $1.47\%$ (relative reduction of $16.9\%$). The benefits extend to low-NFE distilled sampling: ZipVoice-Distill at 4 NFEs improves all three metrics on each benchmark, including a WER decrease from $1.39\%$ to $1.33\%$ on Seed-TTS test-zh (relative reduction of $4.3\%$). 
The following ablations examine block-wise trajectory discrepancies (\S~\ref{subsec:ablation_ot}) and the effect of correction timing (\S~\ref{subsec:abl_interval}).

\begin{table*}[th]
    \caption{Zero-shot TTS results with and without PCI using frozen pretrained checkpoints: $^*$ -- unofficial E2~TTS, provided by the F5-TTS~\cite{chen-etal-2024-f5tts}; $^\dagger$ -- official F5-TTS~\cite{chen-etal-2024-f5tts}; $^\ddagger$ -- official ZipVoice(-Distill)~\cite{zhu2025zipvoice}. Higher SIM-o and UTMOS and lower WER are better.}
    \label{tab:prompt_conditioned_large}
    \centering
    \resizebox{\textwidth}{!}{%
        \begin{NiceTabular}{lccccccccccccc}
        \CodeBefore
            \rowcolor{gray!25}{1-2}
        \Body
            \toprule
            \multirow{2}{*}[-3pt]{\textbf{Model}} &
            \multirow{2}{*}[-3pt]{\textbf{Training Data}} &
            \multirow{2}{*}[-3pt]{\textbf{NFEs}} &
            \multicolumn{3}{c}{\textbf{LibriSpeech-PC test-clean}~\cite{10389666_LibriSpeech_PC}} &&
            \multicolumn{3}{c}{\textbf{Seed-TTS test-en}~\cite{commonvoice:2020_Seed_EN}} &&
            \multicolumn{3}{c}{\textbf{Seed-TTS test-zh}~\cite{Guo2020DidispeechAL_Seed_ZH}} \\
            \cmidrule{4-6}
            \cmidrule{8-10}
            \cmidrule{12-14}
            & & &
            \textbf{SIM-o$\uparrow$} &
            \textbf{UTMOS$\uparrow$} &
            \textbf{WER(\%)$\downarrow$} &&
            \textbf{SIM-o$\uparrow$} &
            \textbf{UTMOS$\uparrow$} &
            \textbf{WER(\%)$\downarrow$} &&
            \textbf{SIM-o$\uparrow$} &
            \textbf{UTMOS$\uparrow$} &
            \textbf{WER(\%)$\downarrow$} \\

            \midrule

            Ground Truth & - & -
            & 0.690 & 4.10 & 1.87
            && 0.734 & 3.52 & 2.14
            && 0.755 & 2.78 & 1.25 \\

            \midrule

            E2 TTS$^*$ & LibriHeavy (EN) & 32
            & 0.692 & 3.42 & 2.72
            && 0.706 & 3.21 & 2.32
            && 0.733 & 2.26 & 2.72 \\

            E2 TTS$^*$ \,+\, \textbf{PCI} & LibriHeavy (EN) & 32
            & 0.698 & 3.46 & 2.44
            && 0.712 & 3.27 & 2.32
            && 0.742 & 2.42 & 2.58 \\

            \midrule

            F5-TTS$^\dagger$ & Emilia (EN+ZH) & 32
            & 0.655 & 3.85 & 1.89
            && 0.664 & 3.67 & 1.85
            && 0.750 & 2.91 & 1.70 \\

            F5-TTS$^\dagger$ \,+\, \textbf{PCI} & Emilia (EN+ZH) & 32
            & 0.674 & 3.85 & 1.61
            && 0.685 & 3.86 & 1.82
            && 0.765 & 2.91 & 1.63 \\

            \midrule

            ZipVoice$^\ddagger$ & Emilia (EN+ZH) & 16
            & 0.665 & 3.98 & 1.77
            && 0.696 & 3.82 & 1.70
            && 0.750 & 3.15 & 1.45 \\

            ZipVoice$^\ddagger$ \,+\, \textbf{PCI} & Emilia (EN+ZH) & 16
            & 0.676 & 4.09 & 1.47
            && 0.694 & 3.90 & 1.56
            && 0.761 & 3.22 & 1.38 \\

            \midrule

            ZipVoice-Distill$^\ddagger$ & Emilia (EN+ZH) & 8
            & 0.645 & 4.07 & 1.65
            && 0.670 & 3.90 & 1.64
            && 0.739 & 3.11 & 1.37 \\

            ZipVoice-Distill$^\ddagger$ \,+\, \textbf{PCI} & Emilia (EN+ZH) & 8
            & 0.657 & 4.16 & 1.50
            && 0.684 & 3.89 & 1.59
            && 0.746 & 3.16 & 1.32 \\

            \midrule

            ZipVoice-Distill$^\ddagger$ & Emilia (EN+ZH) & 4
            & 0.649 & 4.05 & 1.64
            && 0.679 & 3.82 & 1.59
            && 0.747 & 3.16 & 1.39 \\

            ZipVoice-Distill$^\ddagger$ \,+\, \textbf{PCI} & Emilia (EN+ZH) & 4
            & 0.658 & 4.11 & 1.57
            && 0.684 & 3.93 & 1.51
            && 0.751 & 3.17 & 1.33 \\

            \bottomrule
        \end{NiceTabular}%
    }
    \vspace{-0.15in}
\end{table*}

\subsection{Ablation I: Block-Wise Trajectory Discrepancies}\label{subsec:ablation_ot}

The following ablation examines prompt drift and its relationship to the generated trajectory. We compare baseline sampling and PCI on ZipVoice-Distill(8 NFEs) using LibriSpeech-PC test-clean~\cite{10389666_LibriSpeech_PC}, with a uniform partition of $[0,1]$. For this diagnostic, the generated region is assigned the reference feature length so the two states' dimensions match; this does not imply phonetic alignment. We quantify the prompt drift defined in \eqref{eq:drift} by its Frobenius norm and extend the reference-path discrepancy measure to the generated region:
\begin{equation}
e_{\Rmask}(t_i) = \|\bm{\delta}_i\|_F, \quad e_{\Gmask}(t_i)= \left\|\xt_{t_i}[\Gmask]-\xt_{t_i}^{\star}[\Gmask]\right\|_F
\label{eq:err_norm}
\end{equation}
Both quantities are measured after each Euler step, before any subsequent PCI correction. Thus, for PCI, $e_{\Rmask}(t_i)$ measures the post-step residual characterized in \eqref{eq:pci_post_step}, rather than the zero prompt discrepancy of the corrected input used at the next velocity evaluation. For the generated region, the reference interpolation is constructed using \eqref{eq:linear_path}. Reference feature values $\bm{f}[\Gmask]$ are used only to compute this diagnostic and are not supplied as sampler inputs; only their sequence length is used to set the generated-region length. Since the learned ODE need not reproduce this particular interpolation, $e_{\Gmask}(t_i)$ is a diagnostic reference-path distance, not a direct measure of numerical integration error or synthesis quality.

\begin{figure}[t]
  \centering
  \includegraphics[width=0.755\linewidth]{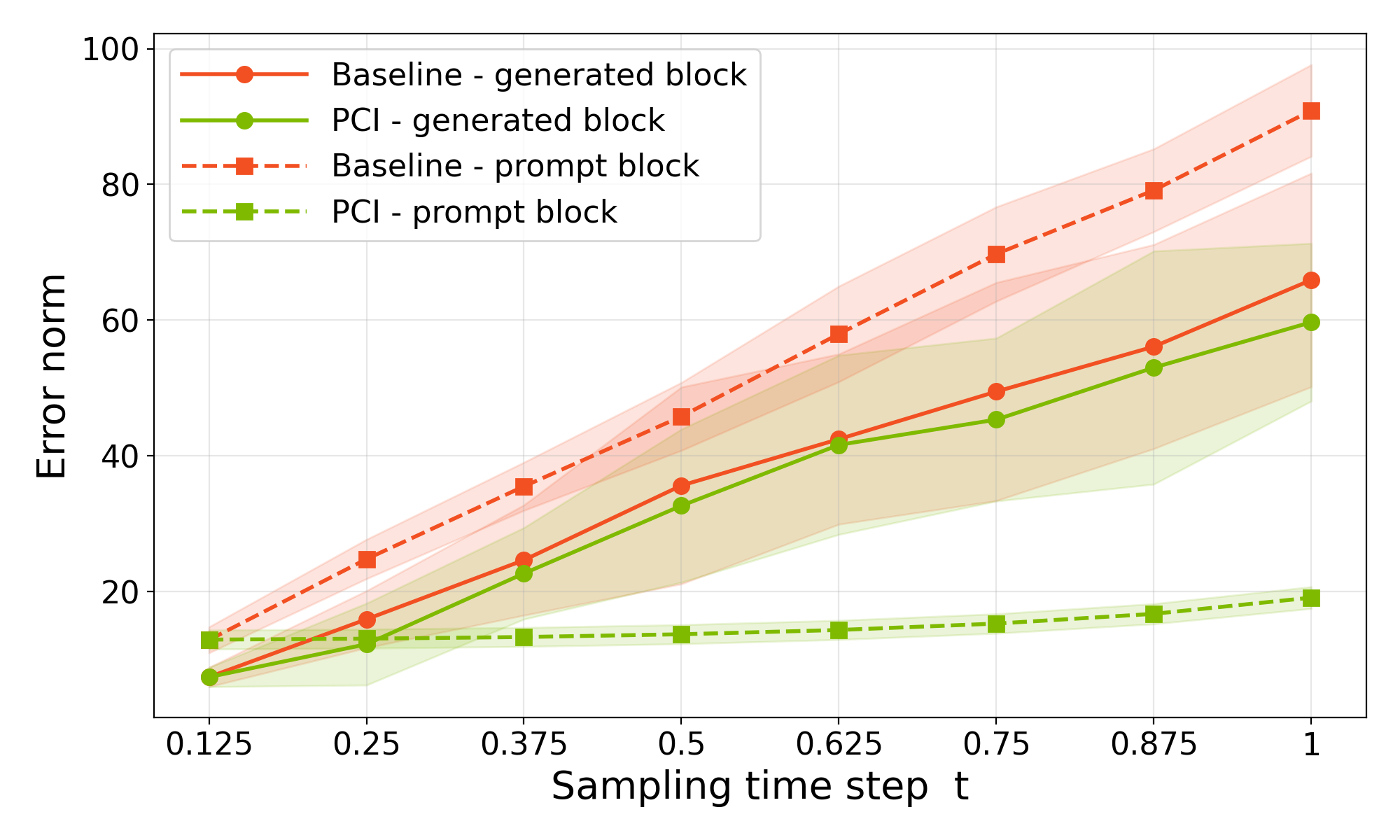}
  \vspace{-0.15in}
  \caption{Prompt drift and generated-block reference-path discrepancy for ZipVoice-Distill (8 NFEs) on LibriSpeech-PC test-clean, with and without PCI. Measurements are taken after each Euler step, before any subsequent prompt correction. Curves show means, with shaded bands indicating one standard deviation across utterances.}
  \label{fig:drift_1}
  \vspace{-0.15in}
\end{figure}

Figure~\ref{fig:drift_1} shows two effects. First, the baseline prompt drift (dashed red) increases along the sampling trajectory, whereas PCI (dashed green) maintains a smaller post-step residual. This behavior is consistent with the distinction in \eqref{eq:drift_recursion} and \eqref{eq:pci_post_step}: baseline sampling carries accumulated prompt drift into subsequent evaluations, while PCI removes it from the network input before the next evaluation. Second, the generated-block discrepancy under PCI (solid green) is also lower than the baseline (solid red), even though PCI never directly overwrites this region. This observation is consistent with prompt-state discrepancies influencing generation through the jointly evaluated velocity field. Together with the output metrics in Table~\ref{tab:prompt_conditioned_large}, these results support the benefit of enforcing prompt consistency beyond the corrected region itself.

\subsection{Ablation II: Correction Windows}\label{subsec:abl_interval}

We next examine which sampling intervals require prompt correction to recover the similarity gains of full PCI. Using the same experimental setup as in \S~\ref{subsec:ablation_ot}, we restrict prompt correction to velocity evaluations within the window $[a,b)$, where $0\leq a\leq b\leq 1$. The corresponding evaluation indices are $\mathcal{I}_{a,b}=\{\,i\in\{0,\ldots,N-1\}:a\leq t_i<b\,\}$. The network input for windowed PCI is defined as
\begin{equation}
\xt_{t_i}^{\mathrm{PCI}}[\Gmask]= \xt_{t_i}[\Gmask], \quad 
\xt_{t_i}^{\mathrm{PCI}}[\Rmask]=
    \begin{cases}
        \bm{p}_{t_i}^{\star}, & i\in\mathcal{I}_{a,b},\\
        \xt_{t_i}[\Rmask], & \text{otherwise}.
    \end{cases}
\label{eq:windowed_pci}
\end{equation}
The Euler update then proceeds as in \eqref{eq:pci}; outside the correction window, the raw state is used without prompt correction as in \eqref{eq:euler}. Setting $a=0$ and $b=1$ applies PCI correction at every step of ODE solver, recovering full PCI. When $a=b$, we apply no correction ($\mathcal{I}_{a,a}=\varnothing$), recovering baseline sampling.

\begin{figure}[t]
  \centering
  \includegraphics[width=0.55\linewidth]{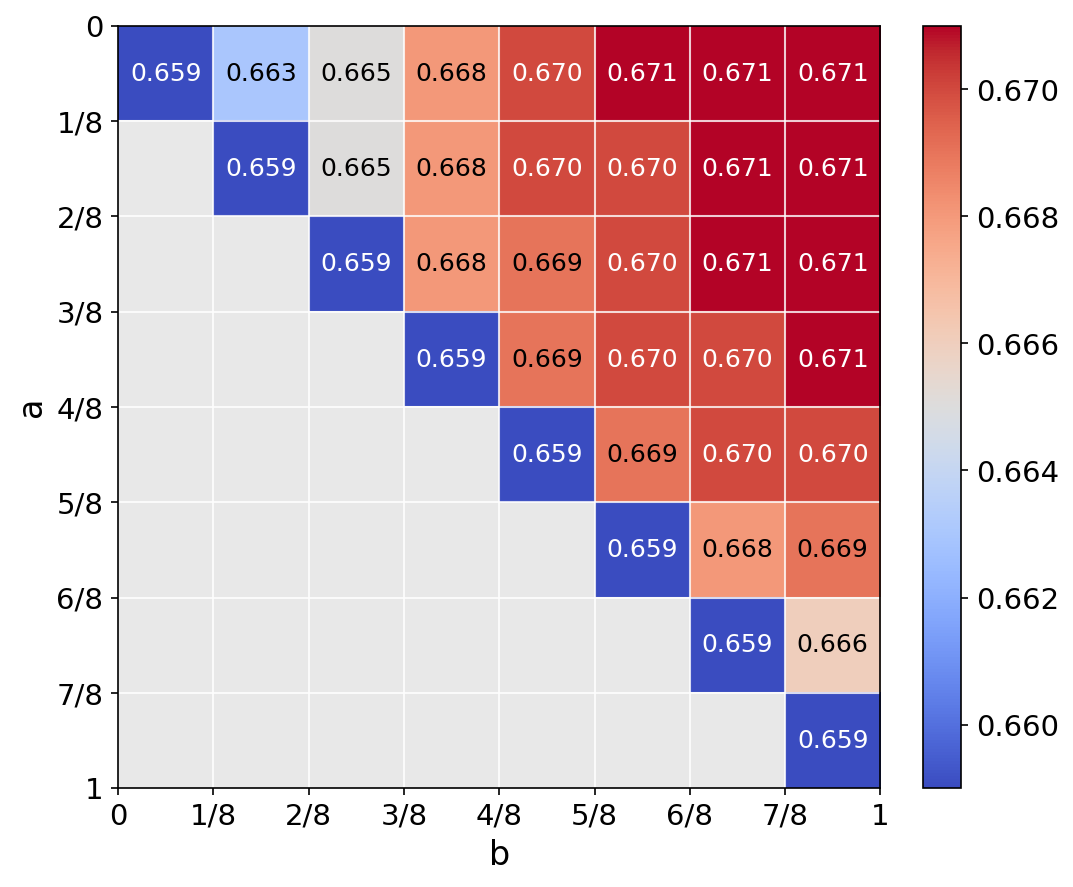}
  \vspace{-0.15in}
  \caption{SIM-o for windowed PCI on ZipVoice-Distill (8 NFEs) on LibriSpeech-PC test-clean. Rows specify window start $a$ and columns -- its end $b$. The top-right entry shows full PCI; diagonal entries correspond to baseline.}
  \label{fig:heatmap}
  \vspace{-0.15in}
\end{figure}

Figure~\ref{fig:heatmap} reports mean SIM-o across utterances, with the window start $a$ on the vertical axis and the window end $b$ on the horizontal axis. Full PCI occupies the top-right corner and achieves the highest similarity, while the diagonal represents baseline sampling. Similarity generally increases as $b$ approaches $1$, and correction over the second half of the schedule captures most of the benefit. Together with the trajectory analysis in \S~\ref{subsec:ablation_ot}, these results suggest that maintaining prompt consistency during later velocity evaluations is particularly important for speaker similarity.

\section{Conclusion}
\label{sec:conclusion}

We introduced Prompt-Consistency Inference (PCI), an inference-time correction that restores the prompt state to its prescribed conditional path before each velocity evaluation. PCI improves speaker similarity and intelligibility across the evaluated models and benchmarks, while generally preserving or improving predicted naturalness, without additional network evaluations. Our ablations show that PCI limits post-step prompt drift and generated-block reference-path discrepancies, despite leaving that block unmodified. Windowed correction further indicates that later sampling intervals are particularly important for recovering speaker similarity gains. These findings highlight prompt consistency as an efficient way to improve flow-based zero-shot speech synthesis.


\clearpage
\section*{Compliance with Ethical Standards}

This study used existing speech benchmarks and did not recruit human participants or conduct animal experiments. No new ethical approval was required for the reported evaluation. Zero-shot voice generation can enable impersonation; reference audio should be used with appropriate permission, and generated speech should be disclosed as synthetic. The authors declare no conflicts of interest.

\bibliographystyle{IEEEbib}
\bibliography{mybib}

\end{document}